\documentclass[prd,twocolumn,aps,amsmath,amssymb,nofootinbib,preprintnumbers]
{revtex4}

\usepackage{graphicx}% Include figure files
\usepackage{dcolumn}% Align table columns on decimal point
\usepackage{bm}% bold math
\usepackage{amsmath}
\usepackage{amsfonts}
\def\ls{\mathrel{\lower4pt\vbox{\lineskip=0pt\baselineskip=0pt
           \hbox{$<$}\hbox{$\sim$}}}}
\def\gs{\mathrel{\lower4pt\vbox{\lineskip=0pt\baselineskip=0pt
           \hbox{$>$}\hbox{$\sim$}}}}
\def\drawbox#1#2{\hrule height#2pt

\hbox{\vrule width#2pt height#1pt \kern#1pt
              \vrule width#2pt}
              \hrule height#2pt}

\def\Asym#1#2{\vcenter{\vbox{\drawbox{#1}{#2}
              \kern-#2pt       % line up boxes
              \drawbox{#1}{#2}}}}

\newcommand{\beq}{\begin{equation}}
\newcommand{\eeq}{\end{equation}}

\begin{document}

%
%\vspace*{2cm}
\title{Probing the parameter space for an MSSM inflation and
the neutralino dark matter}

\author{Rouzbeh Allahverdi$^{1}$}
\author{Bhaskar Dutta$^{2}$}
\author{Anupam Mazumdar$^{3}$}

\affiliation{$^{1}$~Perimeter Institute for Theoretical Physics, Waterloo, ON,
N2L 2Y5, Canada \\ $^{2}$~Department of Physics, Texas A\&M
University, College Station, TX 77843-4242, USA\\ $^{3}$~NORDITA,
Blegdamsvej-17, Copenhagen-2100, Denmark}

%\date{May 3, 2006}

\begin{abstract}
The flat directions $LLe$ and $udd$ within the minimal supersymmetric
Standard Model provide all the necessary ingredients for a successful
inflation with the right amplitude of the scalar density
perturbations, negligible gravity waves and the spectral tilt within
$2\sigma$ observed range $0.92 \leq n_s \leq 1.0$. In this paper we
explore the available parameter space for inflation in conjunction
with a thermal cold dark matter abundance within the minimal
supergravity model. Remarkably for the inflaton, which is a
combination of squarks and sleptons, there is a stau-neutralino
coannihilation region below the inflaton mass $500$~GeV for the
observed density perturbations and the tilt of the spectrum.  For such
a low mass of the inflaton the LHC is capable of discovering the
inflaton candidates within a short period of its operation.  Inflation
is also compatible with the focus point region which opens up for the
inflaton masses above TeV. We show that embedding MSSM within $SO(10)$
can naturally favor this region.\\

MIFP-07-06, February, 2007
\end{abstract}

\maketitle

%%%%%%%%%%%%%%%%%%%%%%%%%%%%%%%%%%%%%%%%%%%%%%%%%%%%%%%%%%%%%%%%%%%%%%%%%%%%%%%
%%%%%%%%%%%%%%%%%%%%%%%%%%%%%%%%%%%%%%%%%%%%%%%%%%%%%%%%%%%%%%%%%%%%%%%%%%%%%%%
\section{Introduction}

The Minimal Supersymmetric Standard Model (MSSM) provides us all the
necessary ingredients for a successful slow roll inflation which can
explain the flatness, homogeneity and the isotropy problems of the
hot big bang cosmology~\cite{AEGM,AKM,Maz-Roz,AEGJM,AJM}, for
details see~\cite{AEGJM}~\footnote{Those readers who wish to seek
models of an observable universe without invoking inflation,
see~\cite{BBMS}}. Not only that the inflaton carries the Standard
Model (SM) charges, the candidates are the two flat directions
within MSSM; $LLe$ and $udd$, where $L$ stands for the superfields
corresponding to lepton doublets, while $u,~d$ and $e$ are the right
handed components of (up and down type) squarks and selectron
respectively~\footnote{This is the first realistic example of the
inflaton where it carries the SM charges, all other attempts failed
in the past~\cite{FEW}. Another gauge invariant candidate for the
inflaton, $NH_uL$ with the neutrino $N$ being Dirac, has also been
proposed in Ref.~\cite{AKM}. This model has a similar prediction to
that of $LLe$ and $udd$, but relies solely on the renormalizable
interactions. In Ref.~\cite{AJM} we explored the possibility of a
gauge invariant inflaton within gauge mediated supersymmetry
breaking.}.

Moreover the biggest advantage of having an MSSM inflaton is the
predictivity, the model parameters (the mass and the couplings) will
hopefully be
%constrained in the particle physics model which we are hoping to
discovered at the LHC~\cite{LHC}. The MSSM inflaton is
robust~\footnote{Furthermore the model does not suffer from
supergravity (SUGRA) and trans-Planckian
corrections~\cite{AEGM,AEGJM}. The well known $\eta$ problem is
absent.} in its prediction on the Cosmic Microwave Background (CMB)
radiation; the model yields the right amplitude for the scalar
density perturbations, the scalar spectral index lies within the
allowed range $0.92\leq n_s \leq 1.0$~\cite{WMAP3,KKMR}, which is
within the $2\sigma$ error bar. The model does not predict any
observable gravity waves in concurrence to the CMB observations nor
does it produce any significant non-Gaussianity.

The end of inflation marks the coherent oscillations of the inflaton
and subsequent decay of the flat direction. The decay products are
distinctively MSSM degrees of freedom by virtue of the inflaton's
gauge couplings to the SM quarks/squarks and gauge bosons/gauginos.
The MSSM degrees of freedom reheat the Universe to a thermal bath at a
temperature above the TeV scale~\footnote{The final reheating is
obtained only when all flat directions are completely
evaporated~\cite{AM1}.}. Such a reheat temperature is sufficient to
answer two of the outstanding puzzles; thermal production of Cold Dark
Matter (CDM) and baryon asymmetry via electroweak baryogenesis within
MSSM~\cite{AEGJM}.

It is well known that the Lightest Supersymmetric Particle (LSP) is
absolutely stable and can be a candidate for the CDM. In most models
the lightest neutralino is mostly bino (the superpartner of the
hypercharge gauge boson), with negligible admixture of wino and/or
Higgsino. It is then a natural question to ask what is the overlapping
parameter region of MSSM which would predict a successful inflation
and also thermally generated neutralino~\footnote{In models with gauge
mediated SUSY breaking gravitino is the LSP. In
Ref.~\cite{AJM}, we have studied gravitino dark matter in conjunction
with MSSM inflation in these models.}.

At the Grand Unified (GUT) scale there is a particular ansatz for
the soft supersymmetry (SUSY) breaking parameters which is very well
motivated in the literature known as the 'mSUGRA' (Constrained
MSSM)~\cite{sugra,sugra1}, it assumes that all of the squark and
slepton soft masses are the same at the  GUT scale to suppress
flavor violation. Similarly all of the A-terms are also taken to be
flavor independent and universal at the GUT scale as well. Finally
all of the gaugino masses are taken to be the same at the GUT scale.
With this ansatz, the parameters are RGE evolved to the TeV scale
and masses and interactions of the particles are studied. The recent
dark matter constraints and other experimental results have
separated this parameter space mostly into three basic regions:
stau-neutralino coannihilation, A-annihilation, and focus point
region~\cite{darkrv}.  The most interesting question is whether the
allowed values of $m_{\phi}$ from the inflation constraints fall in
any of these regions. Moreover, it is a burden on any inflationary
model and the underlying theory to provide conditions for the
observable CDM abundance and the candidate~\footnote{ We notice that
the low scale of MSSM inflation makes it difficult to invoke a late
stage of entropy release which can dilute thermal overabundance of
LSP (as it happens in the bulk region), or, produce non-thermal dark
matter.}.

In order to address these questions, we will first consider the
observables and constraints from CMB and then use them to constrain
the inflaton mass. We will then use this inflaton mass to investigate
the mSUGRA parameter space for the allowed neutralino type dark
matter. We will also discuss the flat directions when right-handed
(RH) neutrinos are present in the model and the consequences of
embedding MSSM in $SU(5)$ or $SO(10)$ models of grand unified theory
(GUT).

The paper is organized as follows. In section 2, we review inflation
in MSSM. In section 3, we discuss the constraints on the inflaton
mass arising from spectral index and the amplitude of perturbations.
In section 4, we discuss the mSUGRA parameter space and the
constraints arising from the allowed values of inflaton mass. In
section 5, we discuss the flat directions and their lifting when RH
neutrinos are added, and possible embedding in $SU(5)$ and $SO(10)$.
Section 6 contains our conclusions.

%%%%%%%%%%%%%%%%%%%%%%%%%%%%%%%%%%%%%%%%%%%%%%%%%%%%%%%%%%%%%%%%%%%

\section{A brief review of MSSM inflation}

\subsection{Inflation near a saddle point}

Let us recapitulate the main features of MSSM flat direction
inflation~\cite{AEGM,AEGJM}. In the limit of unbroken SUSY the flat
directions have exactly vanishing potential. This situation changes
when soft SUSY breaking and non-renormalizable superpotential terms of
the type~\cite{KARI-REV}
\beq \label{supot}
W_{non} = \sum_{n>3}{\lambda_n \over n}{\Phi^n \over M^{n-3}}\,,
\eeq
are included. Here $\Phi$ is a {\it gauge invariant} superfield which
contains the flat direction.  Within MSSM all the flat directions are
lifted by non-renormalizable operators with $4\le n\le 9$~\cite{GKM},
where $n$ depends on the flat direction. We expect that quantum
gravity effects yield $M=M_{\rm P}=2.4\times 10^{18}$~GeV and
$\lambda_n\sim {\cal O}(1)$~\cite{DRT}.

Let us focus on the lowest order superpotential term in
Eq.~(\ref{supot}) which lifts the flat direction. Soft SUSY breaking
induces a mass term and an $A$-term so that the scalar potential along
the flat direction reads:
\beq \label{scpot}
V = {1\over2} m^2_\phi\,\phi^2 + A\cos(n \theta  + \theta_A)
{\lambda_{n}\phi^n \over n\,M^{n-3}_{\rm P}} + \lambda^2_n
{{\phi}^{2(n-1)} \over M^{2(n-3)}_{\rm P}}\,,
\eeq
Here $\phi$ and $\theta$ denote respectively the radial and the
angular coordinates of the complex scalar field
$\Phi=\phi\,\exp[i\theta]$, while $\theta_A$ is the phase of the
$A$-term (thus $A$ is a positive quantity with dimension of mass).
The maximum impact from the $A$-term is obtained when $\cos(n \theta +
\theta_A) = -1$ (which occurs for $n$ values of $\theta$).

%Along
%these directions the potential is
%
%\beq \label{scpot2}
%V = {1\over2} m^2_\phi\,\phi^2 -
%A {\lambda_{n}\phi^n \over n\,M^{n-3}_{\rm P}} + \lambda^2_n
%{{\phi}^{2(n-1)} \over M^{2(n-3)}_{\rm P}}\,.
%\eeq
%

In the gravity mediated SUSY breaking case, the $A$-term and the soft
SUSY breaking mass terms are of the same order of magnitude as the
gravitino mass, i.e. $m_{\phi}\sim A \sim m_{3/2}\sim {\cal O}(1)~{\rm
TeV}$. Then, as pointed out in~\cite{AEGM}, if $A$ and $m_\phi$ are
related by
\beq
\label{cond}
A^2 = 8 (n-1) m^2_\phi\,,
\eeq
there is a saddle point:
\beq \label{phi0}
\phi_0 = \left({m_\phi M^{n-3}_{\rm P}\over
\lambda_n\sqrt{2n-2}}\right)^{1/(n-2)}\,.
\eeq
where $V^{\prime}(\phi) = V^{\prime \prime}(\phi_0)=0$.  The potential
is very flat near $\phi_0$, and it is given by:
\beq \label{potential}
V_0 = {(n-2)^2\over2n(n-1)}\,m^2_\phi \phi_0^2\,.
\eeq
As a result, if the flat direction is in the vicinity of $\phi_0$ (and
has a sufficiently small kinetic energy), there will be an ensuing
phase of inflation~\footnote{In Ref.~\cite{AFM} we have addressed how
the flat direction end up at $\phi_0$. This is an initial condition
problem which is addressed if there were prior phases of inflation. In
the context of string theory where there are multiple false vacua
below the string scale, it is conceivable that an eternal inflation is
generic, however, a graceful exit of inflation must require a phase of
MSSM inflation in the observable world to reheat the plasma with the
desired SM degrees of freedom for the Big Bang
Nucleosynthesis~\cite{AEGM,AKM,Maz-Roz,AEGJM,AJM}. In this regard a
string landscape creates an ideal initial condition for the MSSM
inflation.}.

The Hubble expansion rate during inflation is given by
\beq \label{hubble}
H_{\rm inf} = {(n-2) \over \sqrt{6 n (n-1)}}
{m_{\phi} \phi_0 \over M_{\rm P}}\,.
\eeq
Inflation ends when $\vert \eta \vert \sim 1$, where $\epsilon\equiv
(M_{\rm P}^2/2)(V^{\prime}/V)^2$ and $\eta \equiv M^2_{\rm
P}(V^{\prime \prime}/V)$ are the slow roll parameters.  The number of
e-foldings between the time when the observationally relevant
perturbations are generated and the end of inflation follows: ${\cal
N}_{\rm COBE} \simeq 66.9 + (1/4) {\rm ln}({V_0/ M^4_{\rm
P}})$~\cite{MULTI}.  \\

Here we have used the fact that, due to efficient reheating, the
energy density in the inflaton gets converted into MSSM radiation very
quickly after the end of MSSM inflation (for details
see~\cite{AEGJM}). The amplitude of the perturbations thus produced is
given by:
\beq \label{ampl}
\delta_{H} \simeq
\frac{1}{5\pi} \sqrt{\frac{2}{3}n(n-1)}(n-2) ~ \Big({m_\phi M_{\rm P} \over
\phi_0^2}\Big) ~ {\cal N}_{\rm COBE}^2\,.
\eeq
The spectral index for the power spectrum is found to be~\cite{AEGM}:
\beq \label{tilt}
n_s = 1 - {4\over {\cal N}_{\rm COBE}}\,.
\eeq
For weak scale supersymmetry, acceptable $\delta_H$ and $n_s$ are
obtained if $n=6$. In this case, see Eqs.~(\ref{phi0},\ref{hubble}),
we will have $\phi_0 \sim 10^{14}$ GeV and $H_{\rm inf} \sim {\cal
O}(1~{\rm GeV})$. This singles out two flat directions as the inflaton
candidate: $LLe$ and $udd$. From Eq.~(\ref{potential}) it turns out
that ${\cal N}_{\rm COBE} \sim 50$, implying that $n_s \simeq
0.92$. This is compatible within the fit from combined WMAP-$3$ and
SDSS data within $2 \sigma$~\cite{WMAP3,KKMR}, though it lies at the
lower end.

%%%%%%%%%%%%%%%%%%%%%%%%%%%%%%%%%%%%%%%%%%%%%%%%%%%%%%%%%%%%%%%%%%%%%%%

\subsection{Deviation from saddle point}

Inflation can still happen for small deviations from the saddle point
condition Eq.~(\ref{cond}). To quantify this, we define a parameter
$\alpha^2$ such that~\cite{AEGJM,Maz-Roz}:
\beq \label{dev}
{A^2 \over 8 (n-1) m^2_{\phi}} \equiv 1 + \Big({n-2 \over 2}\Big)^2
\alpha^2\,.
\eeq
For $\alpha^2 \neq 0$, the saddle point becomes a point of inflection where
$V^{\prime \prime}(\phi_0) = 0$, and
\beq \label{1st}
V^{\prime}(\phi_0) = \Big({n-2 \over 2}\Big)^2 \alpha^2 m^2_{\phi}
\phi_0.
\eeq
If $\alpha^2 < 0$, the potential has a local minimum and a maximum.
In this case the flat direction is trapped in the local minimum. It
will eventually tunnel past the maximum and a period of slow roll
inflation will follow~\cite{AEGJM}. If $\alpha^2 > 0$, the potential
has no maximum or local minimum, and then slow roll inflation occurs
around $\phi_0$.

For $\alpha^2 \neq 0$ the expressions for $n_s$ and $\delta_H$ are
modified as~\cite{LYTH1} (see also~\cite{Maz-Roz})
\beq \label{ampl2} \delta_H = {1 \over 5 \pi} \sqrt{{2 \over 3}
n(n-1)} (n-2) {m_{\phi} M_{\rm P} \over \phi^2_0}{1 \over \Delta^2}
~ {\rm sin}^2 [{\cal N}_{\rm COBE}\sqrt{\Delta^2}]\,, \eeq
and
\beq \label{tilt2} n_s = 1 - 4 \sqrt{\Delta^2} ~ {\rm cot} [{\cal
N}_{\rm COBE}\sqrt{\Delta^2}],, \eeq
where
\beq \label{Delta} \Delta^2 \equiv n^2 (n-1)^2 \alpha^2 {\cal
N}^2_{\rm COBE} \Big({M_{\rm P} \over \phi_0}\Big)^4\,. \eeq
Note that for for $\alpha^2 = 0$, Eqs.~(\ref{ampl2},\ref{tilt2}) are
reduced to~(\ref{ampl},\ref{tilt}) respectively. For $\alpha^2 < 0$,
the spectral index will be smaller than that in Eq.~(\ref{tilt}),
thus outside the $2 \sigma$ region from observations. The more
interesting case, as pointed out in~\cite{Maz-Roz}, happens for
$\alpha^2 > 0$.  We can in this case get all values within the
allowed range $0.92 \leq n_s \leq 1$~\cite{KKMR} for
\beq \label{Delta2} 0 \leq \Delta^2 \leq {\pi^2 \over 4 {\cal
N}^2_{\rm COBE}}\,. \eeq
%

%%%%%%%%%%%%%%%%%%%%%%%%%%%%%%%%%%%%%%%%%%%%%%%%%%%%%%%%%%%%%%%%%%%%%%%%%%%%%%
%%%%%%%%%%%%%%%%%%%%%%%%%%%%%%%%%%%%%%%%%%%%%%%%%%%%%%%%%%%%%%%%%%%%%%%%%%%%%%

\section{Constraints on the inflaton mass}

The inflaton mass, $m_{\phi}$, is constrained by the experimental
data on the spectral index $n_s$~\cite{WMAP3,KKMR} and
$\delta_H$~\cite{liddle}.

\begin{figure}[t]
%\center
\includegraphics[width=8.5cm]{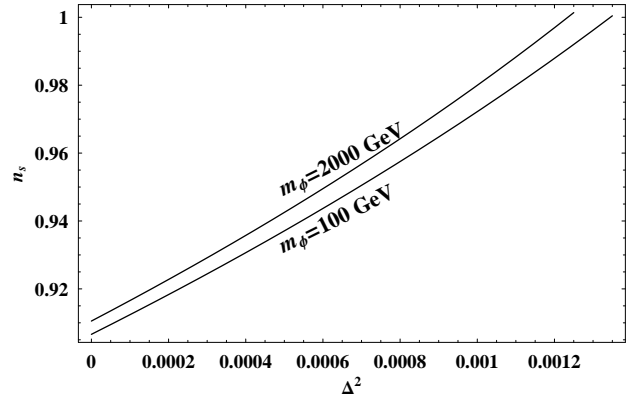}
\caption{$n_s$ is plotted as a function of $\Delta^2$ for different
values of $m_{\phi}$. $\Delta$ is defined in the text.  We choose
$\lambda$ =1.} \label{nsdel0}
\end{figure}

\begin{figure}[t]
\vspace{2cm}
%\center
\includegraphics[width=8.5cm]{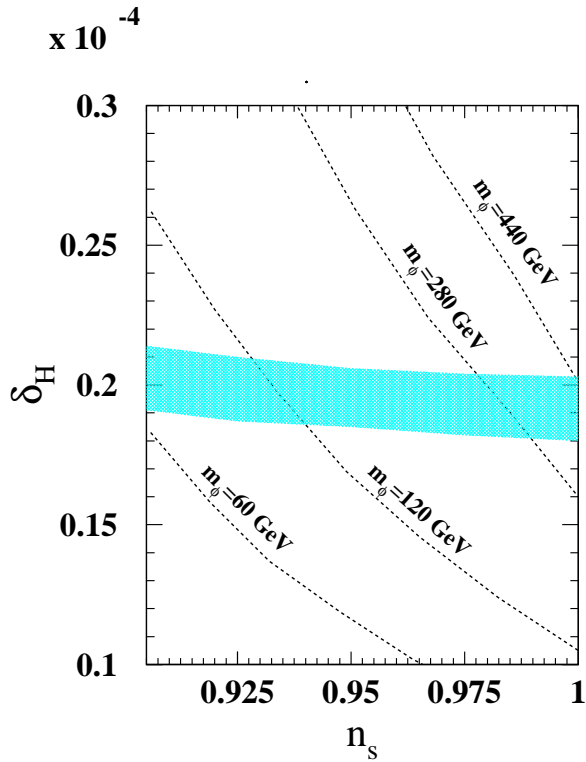}
\caption{$\delta_H$ is plotted as a function of $\Delta^2$ for
different values of $m_{\phi}$. We used $\lambda$ =1. The blue band
denotes the experimentally allowed values of $\delta_H$.}
\label{nsdel}
\end{figure}

We first find the solutions of $m_{\phi}$ by solving
Eqs.~(\ref{ampl2},\ref{tilt2}). $n_s$ depends mainly on $\Delta^2$
and is mostly independent of $m_{\phi}$ and $\lambda$ (the coupling
in Eq.~(1)). The parameter $\Delta^2$ is defined in
Eq.~(\ref{Delta2}). We therefore solve $\Delta^2$ from
Eq.~(\ref{tilt2}) and apply this solution to determine the bounds on
$m_{\phi}$ from the Eq.~(\ref{ampl2}). In figure~\ref{nsdel0}, we
show $n_s$ as a function of $\Delta^2$. The range for $\Delta^2$ is
determined from Eq.~(\ref{Delta2}).\\

In figure~\ref{nsdel}, we show $\delta_H$ as a function of $n_s$ for
different values of $m_{\phi}$. The blue band shows the
experimentally allowed region. We find that smaller values of
$m_{\phi}$ are preferred for smaller values of $n_s$. We also find
that the allowed range of $m_{\phi}$ is $75-440$~GeV for the
experimental ranges of $n_s$ and $\delta_H$.  We assume $\lambda\sim
1$ for these two figures. If $\lambda$ is less than ${\cal O}(1)$,
e.g., $\lambda \sim 0.1$ or so (which can occur in $SO(10)$ model),
it will lead to an increase in $m_{\phi}$. Now we need to study
these allowed ranges of the inflaton mass in the mSUGRA model. Since
the inflaton mass is related to the parameters of the mSUGRA model,
the main question is whether the allowed range of the inflaton mass
is consistent with the experimentally allowed mSUGRA model or not.

%%%%%%%%%%%%%%%%%%%%%%%%%%%%%%%%%%%%%%%%%%%%%%%%%%%%%%%%%%%%%%%%%%%%
%%%%%%%%%%%%%%%%%%%%%%%%%%%%%%%%%%%%%%%%%%%%%%%%%%%%%%%%%%%%%%%%%%%%

\section{Inflation and dark matter in ${\rm m}$SUGRA}

Since $m_{\phi}$ is related to the scalar masses, sleptons ($LLe$
direction) and squarks ($udd$ direction), the bound on $m_{\phi}$
will be translated into the bounds on these scalar masses which are
expressed in terms of the model parameters~\cite{AEGJM}. The models
of mSUGRA depend only on four parameters and one sign. These are
$m_0$ (the universal scalar soft breaking mass at the GUT scale
$M_{\rm G}$); $m_{1/2}$ (the universal gaugino soft breaking mass at
$M_{\rm G}$); $A_0$ (the universal trilinear soft breaking mass at
$M_{\rm G}$)~\footnote{The relationship between the two $A$ terms,
the trilinear, $A_0$ and the non-renormalizable $A$ term in
Eq.(\ref{scpot}) can be related to each other, however, that depends
on the SUSY breaking sector. For a Polonyi model, they are given by:
$A=(3-\sqrt{3})/(6-\sqrt{3})A_0$~\cite{AEGJM}.}; $\tan\beta =
\langle H_2 \rangle \langle H_1 \rangle$ at the electroweak scale
(where $H_2$ gives rise to $u$ quark masses and $H_1$ to $d$ quark
and lepton masses); and the sign of $\mu$, the Higgs mixing
parameter in the superpotential ($W_{\mu} = \mu H_1 H_2$).
Unification of gauge couplings within supersymmetry suggests that
$M_{\rm G} \simeq 2 \times 10^{16}$ GeV. The model parameters are
already significantly constrained by different experimental results.
Most important constraints are:

\begin{itemize}

\item{The light Higgs mass bound of $M_{h^0} > 114.0$~GeV from LEP
\cite{higgs1}.}

\item{The $b \rightarrow s \gamma$ branching ratio~\cite{bsgamma}:
$2.2\times10^{-4} < {\cal B}(B \rightarrow X_s \gamma) <
4.5\times10^{-4}$.}

\item{In mSUGRA the $\tilde\chi^0_1$ is the candidate for CDM.  The
$2\sigma$ bound from the WMAP \cite{WMAP3} gives a relic density bound
for CDM to be $0.095 < \Omega_{\rm CDM} h^2 < 0.129 $.}

\item{The bound on the lightest chargino mass of
$M_{\tilde\chi^{\pm}_1} > 104$~GeV from LEP~\cite{aleph}.}

\item{The possible $3.3~\sigma$ deviation
(using $e^+e^-$ data to calculate the leading order hadronic contribution)from the SM expectation of the
anomalous muon magnetic moment from the muon $g-2$ collaboration
\cite{BNL}.}

\end{itemize}

The allowed mSUGRA parameter space, at present, has mostly three
distinct regions: (i)~the stau-neutralino
($\tilde\tau_1~-~\tilde\chi^1_0$), coannihilation region where
$\tilde\chi^1_0$ is the lightest SUSY particle (LSP), (ii)~the
$\tilde\chi^1_0$ having a dominant Higgsino component (focus point)
and (iii)~the scalar Higgs ($A^0$, $H^0$) annihilation funnel
(2$M_{\tilde\chi^1_0}\simeq M_{A^0,H^0}$). These three regions have
been selected out by the CDM constraint. There stills exists a bulk
region where none of these above properties is observed, but this
region is now very small due to the existence of other experimental
bounds. After considering all these bounds we will show that there
exists an interesting overlap between the constraints from inflation
and the CDM abundance.

We calculate $m_{\phi}$ at $\phi_0$ and $\phi_0$ is $~10^{14}$ GeV
which is two orders of magnitude below the GUT scale. From this
$m_{\phi}$, we determine $m_0$ and $m_{1/2}$ by solving the RGEs for
fixed values of $A_0$ and $\tan\beta$. The RGEs for $m_{\phi}$ are
\begin{eqnarray}
\mu{dm_{\phi}^2\over{d\mu}}&=&{-1\over{6\pi^2}}({3\over
2}{M_2^2}g_2^2+{9\over {10}}{M_1^2}g_1^2)\,,\quad\, ({\rm for\,
LLe})\, \nonumber \\
\mu{dm_{\phi}^2\over{d\mu}}&=&{-1\over{6\pi^2}}({4}{M_3^2}g_3^2+
{2\over {5}}{M_1^2}g_1^2)\,, \quad\, ({\rm for\, udd})\,.
\end{eqnarray}
$M_1,~M_{2}$ and $M_3$ are $U(1),~SU(2)$ and $SU(3)$ gaugino masses
respectively.

After we determine $m_0$ and $m_{1/2}$ from $m_{\phi}$, we can
determine the allowed values of $m_{\phi}$ from the experimental
bounds on the mSUGRA parameters space. In order to obtain the
constraint on the mSUGRA parameter space, we calculate the SUSY
particle masses by solving the RGEs at the weak scale using four
parameters of the mSUGRA model and then use these masses to
calculate Higgs mass, $BR[b\rightarrow s \gamma]$, dark matter
content etc.

\begin{figure}[t]
\vspace{1cm} \center
\includegraphics[width=8.0cm]{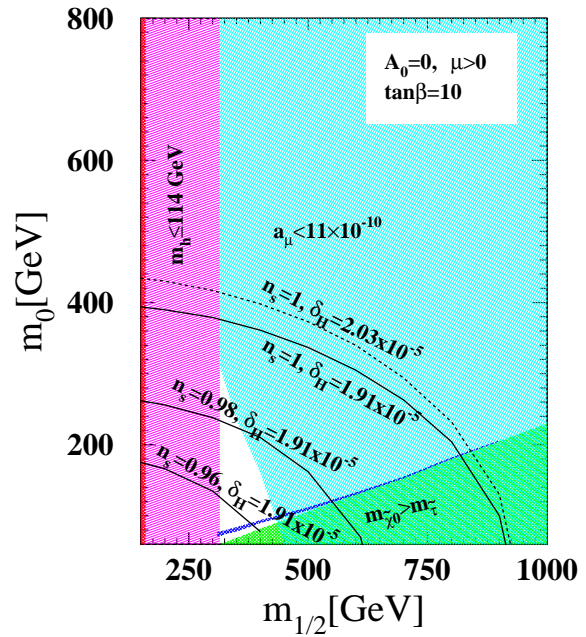} \caption{The
contours for different values of $n_s$ and $\delta_H$ are shown in
the $m_0-m_{1/2}$ plane for $\tan\beta=10$. We used $\lambda=1$ for
the contours. We show the dark matter allowed region {narrow blue
corridor}, (g-2)$_\mu$ region (light blue) for $a_{\mu}\leq
11\times10^{-8}$, Higgs mass $\leq 114$ GeV (pink region) and LEPII
bounds on SUSY masses (red). We also show the the dark matter
detection rate by vertical blue lines.} \label{10flat}
\end{figure}

\begin{figure}[t]
\vspace{1cm} \center \includegraphics[width=8.0cm]{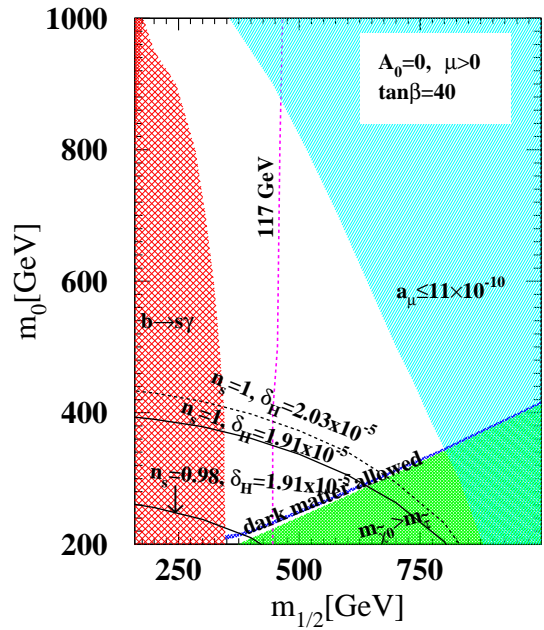}
\caption{The contours for different values of $n_s$ and $\delta_H$
are shown in the $m_0-m_{1/2}$ plane for $\tan\beta=40$. We used
$\lambda=1$ for the contours. We show the dark matter allowed region
{narrow blue corridor}, (g-2)$_\mu$ region (light blue) for
$a_{\mu}\leq 11\times10^{-8}$, $b\rightarrow s\gamma $ allowed
region (brick) and LEPII bounds on SUSY masses (red).}
\label{40flat}
\end{figure}

We show that the mSUGRA parameter space in figures~\ref{10flat},
\ref{40flat} for $\tan\beta=10$ and $40$ with the $udd$ flat
direction using $\lambda=1$~\footnote{We have a similar figure for
the flat direction $LLe$ which we do not show in this paper. All the
figures are for $udd$ flat direction as an inflaton.}. In the
figures, we show contours correspond to $n_s=1$ for the maximum
value of $\delta_H=2.03\times 10^{-5}$ (at $2\sigma$ level) and
$n_s=1.0,~0.98,~0.96$ for $\delta_H=1.91\times 10^{-5}$. The
constraints on the parameter space arising from the inflation
appearing to be consistent with the constraints arising from the
dark matter content of the universe and other experimental results.
 We find that $\tan\beta$ needs to
be smaller to allow for smaller values of $n_s<1$. It is also
interesting to note that the allowed region of $m_{\phi}$, as
required by the inflation data for $\lambda=1$ lies in the
stau-neutralino coannihilation region which requires smaller values
of the SUSY particle masses. The SUSY particles in this parameter
space are, therefore, within the reach of the LHC very quickly. The
detection of the region at the LHC has been considered in
refs~\cite{dka}. From the figures, one can also find that as
$\tan\beta$ increases, the inflation data along with the dark
matter, rare decay and Higgs mass constraint  allow smaller ranges
of $m_{1/2}$. For example, the allowed ranges of  gluino masses are
765 GeV-2.1 TeV and 900 GeV-1.7 TeV for $\tan\beta=10$ and 40
respectively.

\begin{figure}[t]
\vspace{1cm} \center
\includegraphics[width=8.0cm]{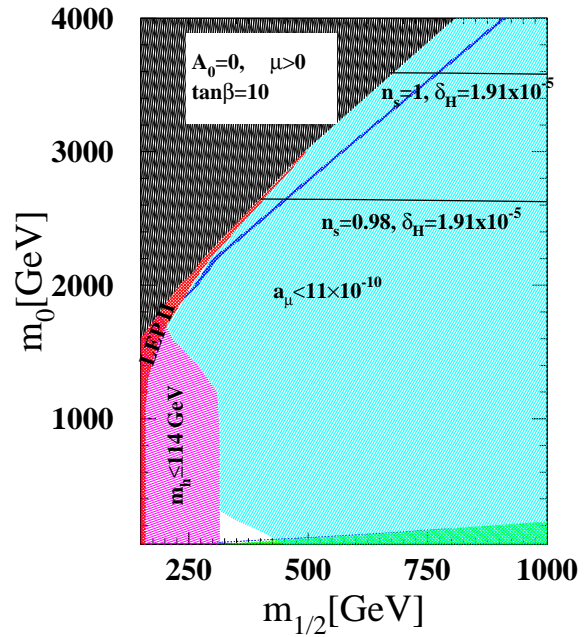}
\caption{The contours for different values of $n_s$ and $\delta_H$
are shown in the $m_0-m_{1/2}$ plane for $\tan\beta=10$. We used
$\lambda=0.1$ for the contours. We show the dark matter allowed
region {narrow blue corridor}, g-2 region (light blue) for
$a_{\mu}\leq 11\times10^{-8}$, Higgs mass $\leq 114$ GeV (pink
region) and LEPII bounds on SUSY masses (red). The black region is
not allowed by radiative electroweak symmetry breaking. We use
$m_t=172.7$~GeV for this graph.} \label{10flatfocus}
\end{figure}

So far we have chosen $\lambda=1$. Now if $\lambda$ is small e.g.,
$\lambda\ls 10^{-1}$, we find that the allowed values of $m_{\phi}$
to be large. In this case the dark matter allowed region requires
the lightest neutralino to have larger Higgsino component in the
mSUGRA model. As we will see shortly, this small value of $\lambda$
is accommodated in $SO(10)$ type model. In figure~\ref{10flatfocus},
we show $n_s=1,\, 0.98$ contours for $\delta_H=1.91\times 10^{-5}$
in the mSUGRA parameter space for $\tan\beta=10$. In this figure, we
find that $n_s$ can not smaller than 0.97, but if we lower $\lambda$
which will demand larger $m_{\phi}$ and therefore $n_s$ can be
lowered down to 0.92.

\begin{figure}[t]
\vspace{1cm}
\includegraphics[width=6.5cm]{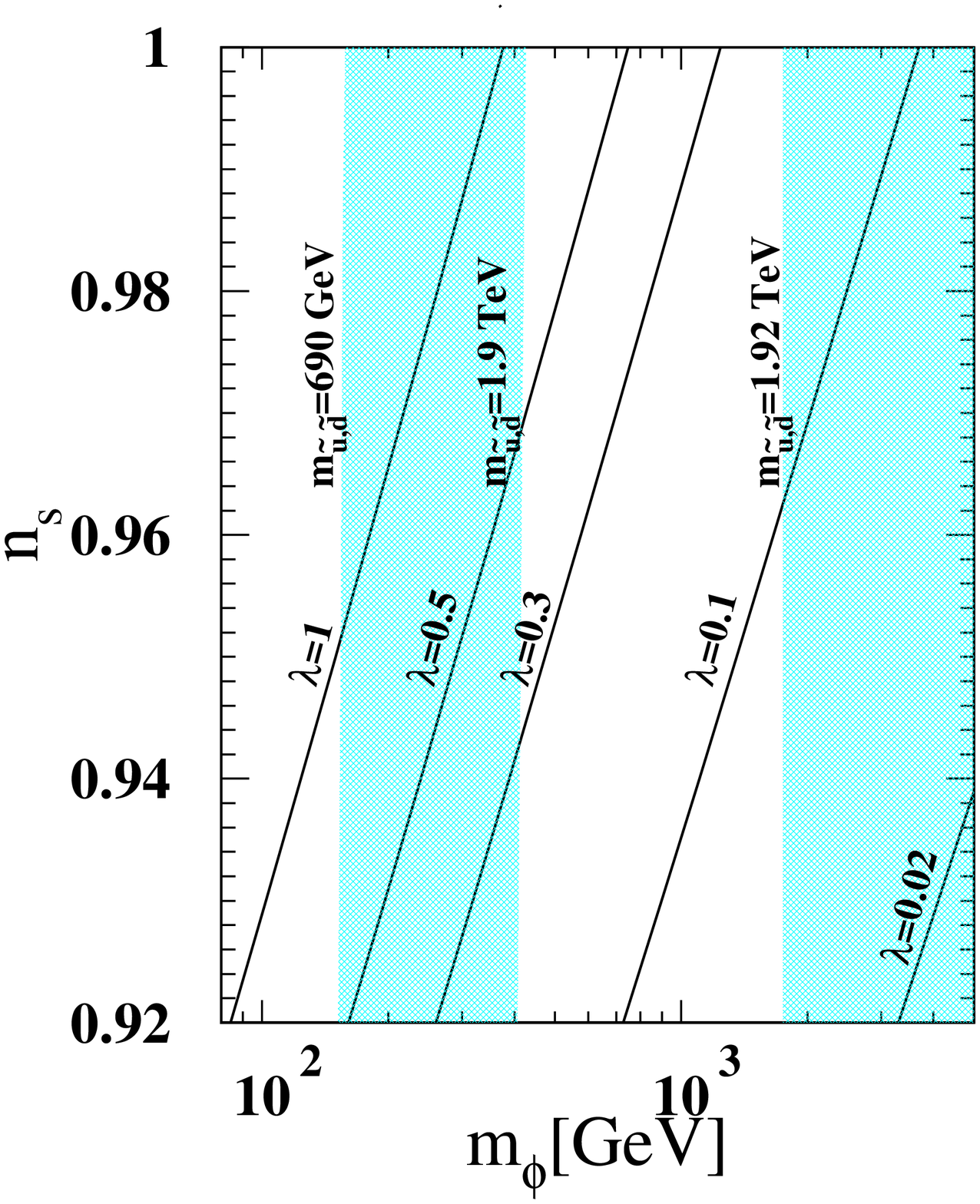}
\caption{Contours of $\lambda$  for $\delta_H=1.91\times 10^{-5}$ in
the $n_s$-$m_{\phi}$ plane. The blue band on the left is due to the
stau-neutralino coannihilation region for $\tan\beta=10$ and the
blue band on the right (which continues beyond the plotting range)
denotes the focus point region.} \label{lamcon}
\end{figure}

In figure~\ref{lamcon}, we show the contours of $\lambda$ for
different values of $m_{\phi}$ which are allowed by $n_s$ and
$\delta_H=1.91\times 10^{-3}$. The blue bands show the dark matter
allowed regions for $\tan\beta=10$. The band on the left is due to
the stau-neutralino coannihilation region allowed by other
constraints and the allowed values of $\lambda$ are 0.3-1.  The
first two generation squarks masses are 690 GeV and 1.9 TeV for  the
minimum and maximum values of $m_{\phi}$ allowed by the dark matter
and other constraints. The gluino masses for these are $765$~GeV and
$2.1$~TeV respectively.  The band is slightly curved due to the
shifting of $\phi_0$ as a function $\lambda$. (We solve for SUSY
parameters from the inflaton mass at $\phi_0$). The band on the
right which continues beyond the plotting range of the
figure~\ref{lamcon} is due to the Higgsino dominated dark matter. We
find that $\lambda$ is mostly $\leq 0.1$ in this region and
$m_{\phi}>1.9$ TeV. In this case the squark masses are much larger
than the gluino mass since $m_0$ is much larger than $m_{1/2}$.

%%%%%%%%%%%%%%%%%%%%%%%%%%%%%%%%%%%%%%%%%%

%%%%%%%%%%%%%%%%%%%%%%%%%%%%%%%%%%%%%%%%%%%%

%%%%%%%%%%%%%%%%%%%%%%%%%%%%%%%%%%%%%%%%%%%%%%%

\section{Grand unified Models and Inclusion of Right-Handed Neutrinos}

\subsection{Embedding MSSM inflation in $SU(5)$ or $SO(10)$ GUT}

As we have pointed out, mSUGRA makes a mild assumption that there
exists a GUT physics which encompasses MSSM beyond the unification
scale $M_{\rm G}$~\footnote{We remind the readers that inflation
occurs around a flat direction VEV $\phi_0 \sim 10^{14}$ GeV. Since
$\phi_0 \ll M_{\rm G}$, heavy GUT degrees of freedom play no role in
the dynamics of MSSM inflation, and hence they can be ignored.}. Here
we wish to understand how such embedding would affect inflationary
scenario, for instance, would it be possible to single out either
$LLe$ or $udd$ as a candidate for the MSSM inflaton.

The lowest order non-renormalizable superpotential terms which lift
$LLe$ and $udd$ are (see Eq.~(\ref{supot})):
\beq \label{supot2}
{(LLe)^2 \over M^3_{\rm P}} ~ ~ , ~ ~ {(udd)^2 \over M^3_{\rm P}}.
\eeq
It is generically believed that gravity breaks global symmetries.
Then all {\it gauge invariant} terms which are $M_{\rm P}$
suppressed should appear with $\lambda \sim {\cal O}(1)$. Obviously
the above terms in Eq.~(\ref{supot2}) are invariant under the SM.
Once the SM is embedded within a GUT at the scale $M_{\rm G}$, where
gauge couplings are unified, the gauge group will be enlarged. Then
the question arises whether such terms in Eq.~(\ref{supot2}) are
invariant under the GUT gauge group or not. Note that a GUT singlet
is also a singlet under the SM, however, the vice versa is not
correct. To answer this question, let us consider $SU(5)$ and
$SO(10)$ models separately.

\begin{itemize}

\item{{\bf $SU(5)$}:\\
We briefly recollect representations of matter fields in this case:
$L$ and $d$ belong to ${\bf {\bar 5}}$, while $e$ and $u$ belong to
${\bf 10}$ of $SU(5)$ group. Thus under $SU(5)$ the superpotential
terms in Eq.~(\ref{supot2}) read
\beq \label{su5}
{{\bf \bar {5}} \times {\bf \bar {5}}
\times {\bf 10} \times {\bf \bar {5}} \times {\bf \bar {5}} \times {\bf 10}
\over M^3_{\rm P}}.
\eeq
This product clearly includes a $SU(5)$ singlet. Therefore in the case
of $SU(5)$, we expect that $M_{\rm P}$ suppressed terms as in
Eq.~(\ref{supot}) appear with $\lambda \sim {\cal O}(1)$~\footnote{If
we were to obtain the $(LLe)^2$ term by integrating out the heavy
fields of the $SU(5)$ GUT, then $\lambda=0$. This is due to the fact
that $SU(5)$ preserves $B-L$.}.}

\item{{\bf $SO(10)$}:\\
In this case all matter fields of one generation are included in the
spinorial representation ${\bf 16}$ of $SO(10)$. Hence the
superpotential terms in Eq.~(\ref{supot2}) are $[{\bf 16}]^6$ under
$SO(10)$, which does not provide a singlet. A {\it gauge invariant}
operator will be obtained by multiplying with a $126$-plet Higgs.
This implies that in $SO(10)$ the lowest order {\it gauge invariant}
superpotential term with $6$ matter fields arises at $n=7$ level:
\beq \label{so10}
{{\bf 16} \times {\bf 16} \times {\bf 16} \times {\bf 16} \times {\bf 16}
\times {\bf 16} \times {\bf 126}_H \over M^4_{\rm P}}\,.
\eeq
Once ${\bf 126}_H$ acquires a VEV, $S0(10)$ can break down to a lower
ranked subgroup, for instance $SU(5)$. This will induce an effective
$n=6$ non-renormalizable term as in Eq.~(\ref{supot}) with
\beq \label{solam}
\lambda \sim \frac{\langle {\bf 126}_H
\rangle}{M_{\rm P}} \sim \frac{{\cal O}(M_{\rm GUT})}{M_{\rm P}}\,.
\eeq
Hence, in the case of $SO(10)$, we can expect $\lambda \sim {\cal
O}(10^{-2}- 10^{-1})$ depending on the scale where SO(10) gets
broken.}

\end{itemize}

We conclude that embedding MSSM in $SO(10)$ naturally implies
$\lambda \ll 1$. Hence an experimental confirmation of the focus
point region may be considered as an indication for $SO(10)$. More
precise determination of the spectral index $n_s$ from future
experiments (such as PLANCK) can in addition shed light on the scale
of $SO(10)$ breaking. Smaller values of $n_s$ (within the range
$0.92 \leq n_s \leq 1$) point to smaller $\lambda$, as can be seen
from  figure 6. This, according to Eq.~(\ref{solam}), implies a
scale of $SO(10)$ breaking, i.e. $\langle {\bf 126}_H \rangle$,
which is closer to the GUT scale.

Further note that embedding the MSSM within $SO(10)$ also provides
an advantage for obtaining a right handed neutrino.

\subsection{Including Right-Handed Majorana Neutrinos}

Eventually one would need to supplement MSSM with additional
ingredients to explain the tiny neutrino masses. Here we consider the
most popular framework; the see-saw mechanism which invokes MSSM plus
three RH (s)neutrinos $N_1,~N_2,~N_3$ with respective Majorana masses
$M_i$.  By adding new superfields to MSSM, one can write a larger
number of non-renormalizable gauge-invariant terms of the form in
Eq.~(\ref{supot}). As a result, a given flat direction might be lifted
at a a different superpotential level. Then a natural question arises
that whether/how adding new superfields will affect the inflaton
candidates, i.e. $LLe$ and $udd$ flat directions.

Since, $N_i$, $1 \leq i\leq 3$, are SM singlets, we can write the
following $n=4$ superpotential terms:
\beq \label{nsup}
{N_i L L e \over M_{\rm P}} ~ ~ , ~ ~ {N_i u d d \over M_{\rm P}}.
\eeq
Note that these terms are also singlet under $SU(5)$ and $SO(10)$.
In the case of $SU(5)$, the terms in Eq.~(\ref{nsup}) read ${\bf
{\bar 5}} \times {\bf {\bar 5}}\times {\bf 10} \times {\bf 1}$,
which includes a singlet. While in the case of $SO(10)$, since $N$
belongs to the ${\bf 16}$, the terms in Eq.~(\ref{nsup}) read ${\bf
16} \times {\bf 16}\times {\bf 16}\times {\bf 16}$, which includes a
singlet. Hence both terms in Eq.~(\ref{nsup}) are allowed in $SU(5)$
or $SO(10)$ embedding of MSSM as well~\footnote{In the case of
$SO(10)$ one can naturally obtain a right-handed neutrino.}.

We now analyze the case for two flat directions separately.

\begin{itemize}

\item{{\bf $LLe$}:\\
First let us consider the $LLe$ flat direction. Taking into account
of the family indices, there are $5$ independent $D$-flat directions
as such~\cite{GKM}. Within MSSM, there are three directions which
are $F$-flat at the $n=3$ level, one of which survives until $n=6$.
However the term in Eq.~(\ref{nsup}) leads to three additional
$F$-term constraints $F_{N_i} = 0$, which are more than sufficient
to lift the remaining direction at the $n=4$ superpotential
level~\footnote{The gauge invariant $LLe$ direction will survive
until $n=6$ if all $M_i \gg \phi_0$.  However this is not a
phenomenologically viable situation.}.

Generically in this case we would expect $LLe$ to be lifted by a
non-renormalizable operator $n < 6$.}

\item{{\bf $udd$}:\\
Next consider the $udd$ direction. With family indices taken into
account, there are $9$ independent $D$-flat directions as
such~\cite{GKM}. Within MSSM, $3$ directions are lifted by $n=4$
terms $uude/M_{\rm P}$, while the remaining $6$ will be lifted at
the $n=6$ level. Note that the superpotential term in
Eq.~(\ref{nsup}) lead to three $F$-term constraints at the $n=4$
level. Nevertheless, $3$ directions will still survive until $n=6$.}

\end{itemize}

Based on the above analysis, if we include the RH neutrinos, we
conclude that $udd$ direction is a more promising inflaton candidate
than $LLe$. The reason is that the flatness of the former will not
be lifted in the presence of physically motivated right handed
neutrino fields in addition to that of the MSSM fields.

%%%%%%%%%%%%%%%%%%%%%%%%%%%%%%%%%%%%%%%%%%%%%%%%%

\section{Discussion and conclusions}

A successful inflation with the right amplitude of the scalar density
perturbations, negligible gravity waves and the spectral tilt can be
described in the context of MSSM by using the $LLe$ or $udd$ flat
direction as the inflaton. The inflaton mass is constrained from the
spectral index and the amplitude of the scalar perturbation. it can be
expressed in terms of the squark and slepton masses for $udd$ and
$LLe$ directions, respectively. The constraints on the inflaton mass
can then be expressed in terms of the bounds on these masses. These
bounds constrain the parameters of the well motivated mSUGRA model.

The parameters of the mSUGRA model are tightly constrained by the
dark matter results along with the results from the LEP experiments
and the rare decays. After considering all these constraints we have
found that an MSSM inflation with a non-renormalizable coupling
$\lambda \sim {\cal O}(1)$ (as expected in an effective field theory
approach) can be explained in the context of mSUGRA and %, which also
%provides the right CDM abundance in
the stau-neutralino coannihilation region is mostly preferred to
satisfy the dark matter content of the universe. The SUSY masses of
this region are mostly within the reach of the LHC. The maximum
value of the gluino mass that is allowed after we include the
inflation data along with the dark matter constraint is around 2
TeV. We have found that the smaller $\tan\beta$ value allows smaller
spectral index which remains within the $2\sigma$ error of the WMAP
data.

Inflation also allows the Higgsino dominated neutralino dark matter,
as happens in the focus point region. For this one would require the
non-renormalizable coupling to be $\lambda \sim {\cal
O}(10^{-2}-10^{-1})$, which can be naturally obtained by embedding
MSSM in $SO(10)$. Any value of $n_s$ in the experimental allowed
range can be fit by a suitable choice of $\lambda$. More precise
determination of the scalar spectral index in future experiments can
in this case shed light on the scale
of $SO(10)$ breaking. %We find this a remarkable interplay between
%particle physics and cosmology.

We also found that the most promising inflaton candidate is $udd$.
This is due to the fact that the lowest non-renormalizable operator
which lifts the flat direction remains $n=6$, even if one includes the
Right Handed Majorana neutrinos. On the other hand $LLe$ can be lifted
earlier by $n=4$ superpotential terms.

Thus our analysis provides an example of a Standard Model {\it gauge
invariant} inflaton giving rise to a successful inflation and explains
the neutralino CDM abundance, which is in agreement with the present
cosmological observations. Moreover this is the first example where
the ingredients of a primordial inflation can be put onto test in a
laboratory physics such as in the case of LHC.

%%%%%%%%%%%%%%%%%%%%%%%%%%%%%%%%%%%%%%%%%%%%%%%%%%%%%%%%%%%%%%%%%%%%%%%%%%%%%%
\section{Acknowledgments}

We wish to thank Kari Enqvist, Nicolao Fornengo and Juan Garcia
Bellido for discussions at various stages of this work.  The work of
RA is supported by Perimeter Institute for Theoretical
Physics. Research at Perimeter Institute is supported in part by the
Government of Canada through NSERC and by the province of Ontario
through MRI. AM is partly supported by the European Union through
Marie Curie Research and Training Network ``UNIVERSENET''
(MRTN-CT-035863).

%%%%%%%%%%%%%%%%%%%%%%%%%%%%%%%%%%%%%%%%%%%%%%%%%%%%%%%%%%%%%%%%%%%%%%%%%%%%%%

\end{document}